\documentclass[final,3p,times,twocolumn,sort&compress]{elsarticle} 

\journal{Results in Physics}

\usepackage{amsmath}
\usepackage{bm}
\usepackage{xspace}


\begin{document}

\begin{frontmatter}



\title{Local equilibration of fermions and bosons}


\author{Georg Wolschin\corref{cor}}
\ead{g.wolschin@thphys.uni-heidelberg.de}

\address{Institut f{\"ur} Theoretische Physik der Universit{\"a}t Heidelberg, Philosophenweg 12-16, D-69120 Heidelberg, Germany, EU}


\begin{abstract}
Local kinetic equilibration is a prerequisite for hydrodynamics to be valid.
Here it is described through a 
nonlinear diffusion equation for finite systems 
of fermions and bosons. The model is solved exactly for constant transport coefficients in both cases.
It has the proper Fermi-Dirac and Bose-Einstein equilibrium limits and can replace the relaxation-time approximation (RTA).
The microscopic transport coefficients are determined through
the macroscopic variables temperature and local equilibration time. 
Applications to the transverse energy of quarks and gluons in the initial stages of central
relativistic heavy-ion collisions, and to bosonic and fermionic atoms at
low energies appropriate for cold quantum gases are discussed.
\end{abstract}

\begin{keyword}
Kinetic equilibration  \sep Nonlinear models  \sep Relativistic heavy-ion collisions \sep Cold quantum gases
\PACS 05.30.-d \sep 05.30.Jp \sep 05.30.Fk 

\end{keyword}

\end{frontmatter}

\section{Introduction}
The evolution of a physical system towards local statistical equilibrium is of general interest. In the realm of quantum physics, fermions and bosons with their respective quantum-statistical properties must be considered separately. Due to the antisymmetry of their states, fermions obey Pauli's principle. Changes of the occupation probabilities of single-particle states are correspondingly suppressed, causing substantially larger local equilibration times than in the case of bosons.
Bosons are not only free to occupy any state, but can also form a condensate at zero momentum, which can have a significant effect on the statistical equilibration process \cite{svi91,setk95,sto97,kas97,dal99} and hence, the final state can deviate considerably from a purely thermal distribution in the infrared.

The full many-body problem for fermions or bosons with mean-field and collision terms at low or high energy can only be solved numerically. Approximate solutions may be obtained with simplified forms of the collision term. As an example, a phenomenological collision term based on a linear relaxation-time approximation (RTA) that governs the equilibration towards the local mean momentum had been added
to the Wigner transform of the one-body density in nonrelativistic calculations \cite{rbw79}.  A related ansatz has been employed in relativistic calculations, such as in Refs.\,\cite{bay84,jpbli18}, and in coupled equations for fermions and bosons\,\cite{fmr18}. 
This does, however, not properly account for the system's intrinsic nonlinearity that is imposed by the dynamical effect of two-body collisions. 

It was therefore proposed in Refs.\,\cite{gw82,bgw18} to replace the relaxation-time approximation for the fermionic collision term by a nonlinear diffusion equation in momentum space which can be solved analytically. A related model for bosons was recently devised in Ref.\,\cite{gw18}, and applied to cold quantum gases in Ref.\,\cite{gw18a}. This work offers a compact and unified comparative treatment of both, fermions and bosons, at high and low energies. 

In the next section, the model is formulated for both cases, and in Sec.\,3 it is solved exactly for constant transport coefficients. Although this represents a highly simplified situation, the correct Fermi-Dirac and Bose-Einstein equilibrium limits are recovered, and the analytical time evolution towards these limits is properly described. Specific features such as antiparticle production of fermions, and condensate formation of bosons are discussed in Sec.\,4. The determination of the microscopic transport coefficients from the macroscopic variables temperature and local equilibration time is performed in Sec.\,5, where the differences between fermionic and bosonic local equilibration times are emphasized. Corresponding calculations for relativistic heavy-ion collisions, and for cold quantum gases are presented. The conclusions are drawn in Sec.\,6.
\section{Kinetic model}
Assuming spatial homogeneity for the distribution function $f\,(x,p,{t})$ with momentum $p$\,
or energy $\epsilon\,(p)$\, and a spherically symmetric momentum dependence, one can reduce
the kinetic equations to one dimension by carrying out the angular integration \cite{fuku17}. 
Hence, in the present approach an exact analytical solution of the local equilibration problem 
is made possible by limiting the consideration to  
a spherical distribution in momentum space during the initial time evolution, 
at least, in the transverse degree of freedom.

Regarding an application to relativistic heavy-ion collisions, the model
is thus tailored to the short initial stages of the collision
only, when local statistical equilibrium in momentum space is achieved for both fermions (quarks) and bosons (gluons)
through collisions. This phase lasts only $\tau_\text{eq}^+ \simeq 0.1$ fm/$c$ for gluons \cite{fmr18} at the
beginning of the creation of the fireball, and it will be shown that it
is about an order of magnitude larger for quarks due to their different statistical properties (Pauli's principle).
In particular, the valence quarks reside in the fragmentation sources \cite{gw16}, and equilibrate locally
on a larger timescale $\tau_\text{eq}^- < 1$ fm/$c$.

Both local equilibration 
processes are best described in transverse momentum space, where 
for central collisions only a single variable is needed, namely,
the transverse energy $\epsilon_\text{T}$.
Given the small rest masses of the up and down quarks that make up the 
projectiles, these can be neglected for the purpose of investigating the local 
equilibration. The mass of the gluons is zero, such that $p_\text{T}\simeq\epsilon_\text{T}$
for both, quarks and gluons.

Local equilibration is a precondition for hydrodynamic approaches \cite{hesne13} 
to be valid. Hydrodynamics then accounts for later phases of the collision lasting typically $5-10$~fm/$c$ at energies
reached at the Relativistic Heavy-Ion Collider (RHIC) or the Large Hadron Collider (LHC). During this later stage,
anisotropic expansion and cooling occurs preferentially in the longitudinal, but also in the transverse direction, 
and eventually highly anisotropic distributions of produced particles are generated. In this work, however, only the
preceding local equilibration stage in the transverse plane is considered, such that isotropy is fulfilled
in central collisions.
For more general non-spherical distributions in coordinate or momentum space, numerical solutions have been discussed, such as in Ref.\,\cite{two82} 
for nuclear collisions at low (MeV-range) energies, and cylindrical symmetry in coordinate space. 
 
With the restriction to spherical symmetry in momentum space, and the dispersion relations $\epsilon\,(p) =\sqrt{ p^2+m^2}$ for relativistic systems and $\epsilon\,(p) = p^2/(2m)$ for nonrelativistic systems, the basic equation for the
single-particle occupation numbers $n_j\equiv\,n_j^{\pm}(\epsilon_j,t)$ for bosons (+) or fermions (-) becomes
\begin{eqnarray}
\frac{\partial n_1^{\pm}}{\partial t}=\sum_{\epsilon_2,\epsilon_3,\epsilon_4}^\infty  \langle V_{1234}^2\rangle\,G\,(\epsilon_1+\epsilon_2, \epsilon_3+\epsilon_4)\times\qquad\\\nonumber 
\bigl[(1\pm n_1)(1\pm n_2)\,n_3\,n_4-
(1\pm n_3)(1\pm n_4)\,n_1\,n_2\bigr]
 \label{boltzmann}
\end{eqnarray}
with the second moment of the interaction $\langle V^2\rangle$ and the energy-conserving function $G$, which has a finite width such that off-shell scatterings between single-particle states that lie apart in energy space become possible. In an infinite system, it becomes a $\delta$-function as in the usual Boltzmann collision term where single-particle energies are time independent,
\begin{equation}
G\,(\epsilon_1+\epsilon_2, \epsilon_3+\epsilon_4)\rightarrow \pi\,\delta\,(\epsilon_1+\epsilon_2-\epsilon_3-\epsilon_4)\,.
\end{equation} 
The collision term can also be written
 in the form of a master
equation with gain and loss terms, respectively
\begin{equation}
\frac{\partial n_1^{\pm}}{\partial t}=(1\pm n_1)\sum_{\epsilon_4} W^\pm_{4\rightarrow1}\,n_4-n_1\sum_{\epsilon_4}W^\pm_{1\rightarrow4}(1\pm n_4) 
 \label{boltz}
\end{equation}
with the transition probability
\begin{equation}
W^\pm_{4\rightarrow1}(\epsilon_1,\epsilon_4,t)=\sum_{\epsilon_2,\epsilon_3}\, \langle V_{1234}^2\rangle\,G\,(\epsilon_1+\epsilon_2, \epsilon_3+\epsilon_4)\,(1\pm n_2)\,n_3
 \label{trans}
\end{equation}
and $W^\pm_{1\rightarrow 4}$ accordingly. The summations are then replaced by integrations, introducing the densities of states $g_j\equiv g(\epsilon_j)$
and $W^\pm_{4\rightarrow 1}=W^\pm_{41}g_1, W^\pm_{1\rightarrow 4}=W^\pm_{14}g_4$,
and $W_{41}=W_{14}$.

An approximation to Eq.\,({\ref{boltz}) can be obtained through a Taylor expansion of $n_4$ and $g_4\,n_4$
around $\epsilon_4=\epsilon_1$ to second order. By introducing transport coefficients via moments of the transition probability
\begin{eqnarray}
D^\pm(\epsilon_1,t)=&\frac{1}{2}\,g_1\int_0^\infty W^\pm(\epsilon_1,x)\,x^2dx\equiv D\,(\epsilon,t),\\
 \varv^\pm(\epsilon_1,t)=&g_1^{-1}\frac{d}{d\epsilon_1}(g_1D^\pm)\equiv \varv\,(\epsilon,t)
 \label{moments}
\end{eqnarray}
one arrives at a nonlinear partial differential equation for $n^{\pm}\equiv n^{\pm}(\epsilon,t)\equiv n^{\pm}(\epsilon_1,t)\equiv n$
 \begin{equation}
\frac{\partial n^{\pm}}{\partial t}=-\frac{\partial}{\partial\epsilon}\left[\varv\,n\,(1\pm n)\mp n^2\frac{\partial D}{\partial \epsilon}\right]+\frac{\partial^2}{\partial\epsilon^2}\bigl[D\,n\bigr]\,.
 \label{bosfereq}
\end{equation}
Dissipative effects are expressed through the drift term $\varv\,(\epsilon,t)$, diffusive effects through the diffusion term $D\,(\epsilon,t)$.
These transport coefficients depend on the strength of the residual two-body interactions, govern the time evolution of the system, and yield a local equilibration that is much faster \cite{gw82,gw18} than what is obtained with a corresponding linear relaxation-time approximation. 

Taking full account of the nonlinearities, and the dependences of $D$ and $\varv$ on energy and time requires a detailed investigation of the energy conserving function and the microscopic interactions between the particles, and would lead to a highly nonlinear and possibly non-Markovian diffusion equation which can only be solved numerically. It would therefore become difficult to link the transport coefficients and thus the strength of the residual interactions to the macroscopic variables temperature and local equilibration time. Nevertheless, a microscopic calculation of $D$ and $\varv$ from the moments equations
is very desirable.

In the limit of constant transport coefficients, the nonlinear bosonic(+)/fermionic(-) diffusion equation  
for the occupation-number distribution $n^{\pm}(\epsilon,t)\equiv n$
becomes
\begin{equation}
\frac{\partial n^{\pm}}{\partial t}=-\varv\,\frac{\partial}{\partial\epsilon}\left[n\,(1\pm n)\right]+D\,\frac{\partial^2n}{\partial\epsilon^2}\,,
 \label{bosfer}
\end{equation}
and this kinetic equation can be solved exactly.
The thermal equilibrium distribution is a stationary solution 
\begin{equation}
n_\text{eq}^{\pm}(\epsilon)=\frac{1}{e^{(\epsilon-\mu^{\pm})/T}\mp1}
 \label{Bose-Einstein}
\end{equation}
with the chemical potential $\mu^{+}<0$ in a finite Bose system, and $\mu^{-}>0$ for fermions.
The equilibrium temperature is given in terms of the transport coefficients, $T=-D/\varv$ with $\varv<0$ since the drift is towards the infrared region. 

The full equation Eq.\,(\ref{bosfereq}) with self-consistently determined $D$ and $\varv$ should also have the correct equilibrium solutions
 Eq.\,(\ref{Bose-Einstein}) for fermions and bosons, but a mathematical proof is beyond the scope of this work.

\section{Analytical solution}
Analytical solutions for physically meaningful nonlinear partial differential equations are rarely available. A notable exception is the Korteweg-de Vries equation \cite{kdv95}, which is of third order in the (single) spatial variable and has soliton solutions. Another example is Burgers' equation \cite{bur48}, which has the structure of a one-dimensional Navier-Stokes equation without pressure term. It has been used to describe fluid flow and, in particular, shock waves in a viscous fluid, and it can be solved through Hopf's transformation \cite{ho50}.

Considering the diffusion equation Eq.\,(\ref{bosfer}) with constant transport coefficients, the nonlinearity that is due to self-consistently determined 
$D$ and $\varv$ is neglected, but the inherent nonlinearity of the equation itself is treated explicitly. In this case, I have proposed two equivalent analytical solution schemes for both, fermions 
\cite{gw82} and bosons \cite{gw18}: Either start with the nonlinear transformation
\begin{equation}
n^{\pm}(\epsilon,t)=\mp\frac{D}{\varv}\,\partial_\epsilon \ln P\,(\epsilon,t)\,,
\label{nltrafo}
\end{equation}
which reduces Eq.\,(\ref{bosfer}) to a linear Fokker-Planck equation for $P\equiv P\,(\epsilon,t)$
that is readily solvable.
Alternatively, perform the linear transformation
\begin{equation}
n^{\pm}(\epsilon,t)=\pm \frac{1}{2\varv}[w\,(\epsilon,t)-\varv]\,,
\label{lintrafo}
\end{equation}
with $w\equiv w(\epsilon,t)$ obeying Burgers' equation
\begin{equation}
\partial_t w+w\,\partial_\epsilon w=D\,\partial_\epsilon^2 w\,,
\label{burgers}
\end{equation}
which can then be solved analytically. In both cases, one can retransform to $n^{\pm}(\epsilon,t)$ and obtain the exact solution of the nonlinear
bosonic/fermionic diffusion equation Eq.\,(\ref{bosfer}) for $n^{\pm}(\epsilon,t)$ as \cite{gw18}
\begin{equation}
n^{\pm} (\epsilon,t)\, = \, \frac{1}{Z^{\pm}(\epsilon,t)} \; \int_{-\infty}^{\infty} \Big[ \mp\frac{1}{2} \pm \frac{\epsilon-x}{2\varv t} \Big] \, f^{\pm}(x,\epsilon,t) \; dx 
\label{n_2}
\end{equation} 
where $Z^{\pm}(\epsilon,t)$ \cite{bgw18}  is the generalized partition function   
\begin{gather}
Z^{\pm}(\epsilon,t) = \, \int_{-\infty}^{\infty} f^{\pm}(x,\epsilon,t) \, dx \;\;\;\;\; \text{with} 
\notag \\
f^{\pm}(x,\epsilon,t) = \exp \left[-\frac{\varv}{2D}  \Bigl( x \pm 2 \int_0^{x} n_0^{\pm}(y)\,dy \Bigr) - \frac{(\epsilon-x)^2}{4Dt} \right].
\label{f}
\end{gather}
This analytical solution is well-suited to replace the linear relaxation ansatz that has often been used in the literature \cite{rbw79,fmr18,bay84,jpbli18}.
The integrals extend over the full energy space.
This is a consequence of the appearance of the Fokker-Planck equation in the first solution method, and of the heat equation in the second. Confining the integrals to the positive-energy domain would lead to a violation of particle-number conservation. Hence, initial values for the negative-energy region must be specified. 
\section{Fermions and bosons}
For fermions, all negative-energy states in the Dirac sea \cite{pamd30} are initially taken as occupied, 
and hence, the initial conditions are
\begin{equation}
n_0^{-}(\epsilon)=\theta\,(1-\epsilon/\mu)\,.
\label{nif}
\end{equation}
Emerging holes at very high temperatures are interpreted as antiparticles, and particle-number conservation for fermions is fulfilled when both particles and antiparticles are considered \cite{gw18,bgw18},
\begin{eqnarray}
\label{nftot}
N_\text{tot}^-=\int_{-\infty}^\infty n^-(\epsilon,t)\,g\,(\epsilon)\,d\epsilon\\\nonumber
=\int_0^\infty n^-(\epsilon,t)\,g\,(\epsilon)\,d\epsilon\\\nonumber
-\int_{-\infty}^0 [1-n^-(\epsilon,t)]\,g\,(\epsilon)=\text{const}\,,
\end{eqnarray}
with the density of states $g(\epsilon)$: Since particles and antiparticles are produced at the same rate according to Eq.\,(\ref{bosfer}), the effective particle number remains constant. The equilibrium value for the number of produced antiparticles relative to the total fermion number $N_\text{tot}^-$ depends on the ratio of chemical potential and temperature \cite{bgw18}, it is given by $(T/\mu)\,\ln\,(1+e^{-\mu/T})$.  The validity of Eq.\,(\ref{nftot}) has been checked numerically in Ref.\,\cite{gw18} and confirmed analytically in
Ref.\,\cite{bgw18}, where also energy conservation was shown to be fulfilled for constant density of states. 
The boundary conditions are $n^-(-\infty,t)=1,~n^-(+\infty,t)=0$.

For bosons, however, all negative-energy states are initially taken as unoccupied, $n^+_0(x<0)=0$, with boundary conditions $n^+(-\infty,t)=n^+(+\infty,t) = 0$. This will cause diffusion out of the positive-energy domain even at very low energies, such as in case of cold quantum gases. 

The additional requirement is therefore that all particles leaving the positive-energy region assemble in the condensate at $p=0$ -- a principle that can be viewed as a bosonic analog to the Dirac-sea postulate for fermions at negative energies. Hence, the particle number in the positive domain will be depleted with time because particles move from the thermal cloud into the condensate. Obviously, the emergence of phase coherence in the condensate can not be described in this nonlinear diffusion approach, but through overall particle-number conservation, the respective number of bosons in the thermal cloud $N^+(t)$, and in the condensate $N_\text{cond}^+(t)$ can be computed,
\begin{eqnarray}
\label{nbtot}
N_\text{tot}^+=\int_{-\infty}^\infty n^+(\epsilon,t)\,g\,(\epsilon)\,d\epsilon\\
=N_\text{cond}^+(t)+\int_0^\infty n^+(\epsilon,t)\,g\,(\epsilon)\,d\epsilon\,.\nonumber
\end{eqnarray}
\section{Transport coefficients and model calculations}
The transport coefficients $D$ and $\varv$ eventually result from a microscopic theory for the $N$-particle system. In case of constant transport coefficients, they are, however, uniquely related to macroscopic variables, in particular, to the local temperature $T  \equiv -D/\varv$, the local equilibration time $\tau^-_{\text{eq}} \equiv 4D/\varv^2$
for fermions \cite{gw82}, and $\tau^+_{\text{eq}} \equiv 4D/(9\varv^2)$ for bosons \cite{gw18}. Whereas the relation between transport coefficients and temperature arises from the requirement that the thermal equilibrium distributions fulfills Eq.\,(\ref{bosfer}), the local equilibration times can be derived from asymptotic expansions of the error functions occurring in the time-dependent analytical solutions \cite{gw82,gw18}. The local equilibration time plays a similar role as the characteristic time scale $\tau_\text{rel}$ that is used in the relaxation-time approximation (RTA). However, when $\tau_\text{rel}=\tau_\text{eq}$ is taken to model the time evolution in the RTA, it was shown in Refs.\,\cite{gw82,gw18} that local equilibration occurs much slower than in the nonlinear case with constant transport coefficients.

For given values of the macroscopic parameters $T$ and $\tau_\text{eq}$, one can now determine the microscopic transport coefficients for fermions as
\begin{equation}
D^- = \frac{4}{\tau^-_{\text{eq}}} \, T^2 \, , \;\;\;\;\;\;\;\;\; \varv^- = -\frac{4}{\tau^-_{\text{eq}}} \, T \,,
\label{tr1}
\end{equation}
and correspondingly for bosons
\begin{equation}
D^+ = \frac{4}{9\,\tau^+_{\text{eq}}} \, T^2 \, , \;\;\;\;\;\;\;\;\; \varv^+ = -\frac{4}{9\,\tau^+_{\text{eq}}} \, T \,.
\label{tr2}
\end{equation}

Hence, with realistic physical values for the temperature $T$, the local equilibration time $\tau_\text{eq}^\pm$, and the initial distribution $n_0(\epsilon,t)$, one can calculate the exact analytical solutions of the nonlinear diffusion equation Eq.\,(\ref{bosfer}) for both, fermions and bosons. 

Here I use as examples at high energies the occupation-number distribution 
of valence quarks (fermions) and gluons (bosons) as functions of the transverse energy in central relativistic heavy-ion collisions, and at low energies the  distribution in cold quantum gases. For a bosonic gas, a truncated equilibrium distribution that is reminiscent of evaporative cooling is taken as initial condition, whereas for a fermionic gas a $\theta$-function initial condition is taken.

In relativistic heavy-ion collisions such as Pb-Pb at energies
reached at the Large Hadron Collider LHC, the initial local temperature in the centre of the expanding fireball that is created in the collision is about $T\simeq500$\, MeV \cite{hnw17}, substantially above the hadronization temperature $T_\text{H}\simeq 160$\,MeV. The initial state contains not only valence quarks (fermions) of the collision partners which populate the fragmentation sources \cite{mtw09,gw16}, but also abundant gluons residing mostly in the anisotropic fireball. Gluons have a substantially smaller equilibration time than quarks due to the larger color factor, and because of their bosonic properties \cite{gw18}. 
Typical local equilibration times for gluons before the onset of hydrodynamic expansion are of the order of $\tau^+_{\text{eq}}\simeq0.1$\,fm/$c$
\cite{fmr18}. Hence, I use as a local equilibration time for valence quarks 
$\tau^-_{\text{eq}}\gtrsim 9\times\tau^+_{\text{eq}}\simeq 0.9$ fm/$c$. 
The fermion diffusion coefficient is then
$D = 1.1$ GeV$^2$/(fm/$c$), 
and the drift coefficient $\varv = -2.2$ GeV/(fm/$c$). 
\begin{figure}[t!]
\begin{center}
\includegraphics[width=7.6cm]{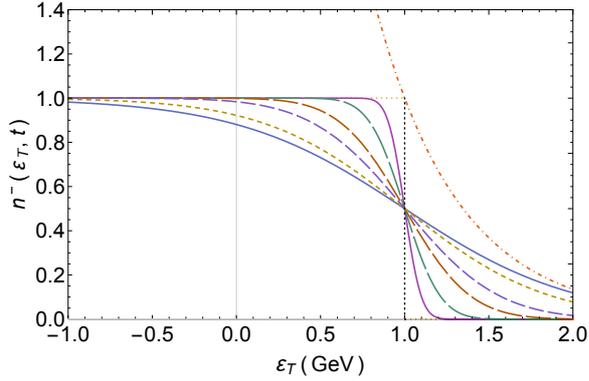}
\caption{\label{fig1} Occupation-number distribution $n^-(\epsilon_\text{\,T},t)$ of a relativistic fermion system (valence quarks) as function of the transverse energy 
$\epsilon_\text{\,T}$ 
including antiparticle production ($\epsilon_\text{\,T}<0)$. It is evaluated analytically
from Eq.\,(\ref{bosfer}) at different times for the initial distribution $n^-_0(\epsilon_\text{\,T}) = \theta\,(1 - \epsilon_\text{\,T}/\mu)$. The parameters are
 $\mu= 1 \, \text{GeV}$, $T = -D/\varv = 500 \, \text{MeV}$, $\tau^-_\text{eq}\,=4D/\varv^2\,= 0.9$ fm/$c$. The distribution 
 $n^-(\epsilon_\text{\,T},t)$ is displayed at 
$ t = 0.003, 0.015, 0.06, 0.15, 0.6$ fm/$c$ (ordered by decreasing dash length towards the solid equilibrium distribution). The upper dot-dashed curve is the Boltzmann distribution.}  
\end{center}
\end{figure}

The result of a corresponding calculation for valence quarks with chemical potential $\mu=\epsilon_\text{F}=1$\,GeV and a $\theta$-function initial distribution is shown in Fig.\,\ref{fig1}. Different from a linear relaxation-time approximation, the analytical solutions of the nonlinear Eq.\,(\ref{bosfer}) are continuous at $\epsilon=\epsilon_\text{\,F}$. They are point-symmetric around the Fermi energy $\epsilon_\text{\,F}$ at all times \cite{bgw18}, and approach the Fermi-Dirac limit for $t\rightarrow\infty$ as expected. As compared to a relaxation ansatz with the same 
$\tau^-_{\text{eq}}$ taken as the relaxation time, the nonlinear local equilibration occurs much faster \cite{gw82,gw18}.
\begin{figure}[t!]
\begin{center}
\includegraphics[width=7.6cm]{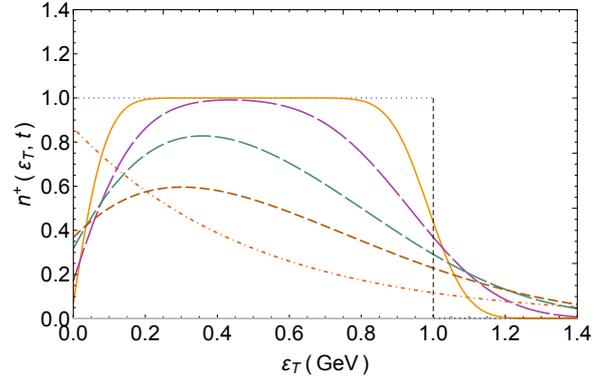}
\caption{\label{fig2} Occupation-number distribution $n^+(\epsilon_\text{\,T},t)$ of a relativistic boson system (gluons) with transverse energy 
$\epsilon_\text{\,T}$  
 evaluated analytically 
from Eq.\,(\ref{bosfer}) at different times for the initial distribution $n_0(\epsilon_\text{\,T}) = \theta\,(1 - \epsilon_\text{\,T}/q_\text{s})$, see text. The parameters are
 $q_\text{s}= 1 \, \text{GeV}$, $\mu=-0.07$ GeV, $T = -D/\varv = 500 \, \text{MeV}$, $\tau^-_\text{eq}\,=4D/(9\varv^2)\,= 0.1$ fm/$c$. The occupation-number distribution 
 $n^+(\epsilon_\text{\,T},t)$ is displayed at 
$ t = 0.003, 0.015, 0.045, 0.09$ fm/$c$ (ordered by decreasing dash length). 
The dot-dashed curve is Boltzmann's distribution, $n_\text{B}(\epsilon)=\exp\,[-(\epsilon-\mu)/T]$.}  
\end{center}
\end{figure}
For sufficiently large times $(t\gtrsim 0.06$ fm/$c$ in Fig.\,\ref{fig1}), the occupation at negative energies falls below one, corresponding to the creation of antiparticle-particle pairs from the filled Dirac sea. In the example shown in Fig.\,\ref{fig1}, the number of created antiquarks is in the equilibrium limit 
$\approx 6.3$\,\% of  the total fermion number $N_\text{tot}^-$.

The result for the local equilibration of gluons with $\epsilon(p)=p$ in the transverse degree of freedom of a central relativistic heavy-ion collision is shown in Fig.\,\ref{fig2}. Here I use again constant transport coefficients and display the analytical solutions of Eq.\,(\ref{bosfer}). In the infrared, the gluon occupation numbers are seen to be depleted due to condensate (BEC) formation at $p=0$. The possible BEC formation in relativistic collisions had been discussed e.g. in Refs. \cite{jpb12,jpb13,jpb14}:
Soft elastic scatterings tend to drive the system towards the formation of a condensate that contains
a large fraction of the gluons while contributing little to the energy density.
In a realistic heavy-ion collision this is, however, unlikely to happen due to inelastic collisions and nonconservation of particle number.
Indeed, it has been found in Ref.\,\cite{blmt16} that at small momenta inelastic collisions dominate the dynamics, and hinder condensation.
Their role for the local equilibration of the system has been emphasized in Ref.\,\cite{ba01},
 and effective kinetic equations in weakly coupled  non-abelian plasmas have been solved numerically in Ref.\,\cite{kur14}.


For the equilibration of a low-energy bosonic system based on Eq.\,(\ref{bosfer}), a cold quantum gas \cite{an95,hul95,ket95} such as $^{87}$Rb is considered, see Fig.\,\ref{fig3}. The relation between energy and momentum arises from the nonrelativistic dispersion relation $\epsilon=p^2/(2m)$. The temperature is chosen as 8 peV$\simeq$\,93 nK, which is below the critical temperature for the formation of a Bose-Einstein condensate. Local equilibration times for the thermal cloud are in the millisecond range; I choose $\tau^+_{\text{eq}}$\,= 3.6 ms. According to Eqs.\,(\ref{tr2}), 
the boson diffusion coefficient becomes
$D = 8\times10^{-3}$\,neV$^2$s$^{-1}$, and the drift coefficient $\varv = -1\,$neV\,s$^{-1} = - 1\,$peV/ms. 
For the initial values $n_0^+(\epsilon)$, a thermal distribution with $\mu= -0.69$\,peV that is truncated at 
$\epsilon_\text{\,c}$\,=\,7\,peV is taken,
\begin{equation}
n_\text{0}^+(\epsilon)=\frac{1}{e^{(\epsilon-\mu)/T}-1}\, \theta\,(1-\epsilon/\epsilon_\text{\,c})\, \theta\,(\epsilon)\,. 
 \label{inibec}
\end{equation}
This is reminiscent of evaporative cooling, where atoms in the high-energy tail are removed to decrease the temperature \cite{an95,lrw96}. It is evident from Fig.\,\ref{fig3} that the time evolution imposed by Eq.\,(\ref{bosfer}) quickly -- within the local equilibration time $\tau^+_{\text{eq}}$ --  produces a new thermal tail in the ultraviolet region that corresponds to a lower temperature of $T\approx 6$\,peV\,$\simeq 70$\,nK. Simultaneously, occupation in the infrared is depleted because atoms disappear from the thermal cloud and condense at $p=0$, keeping the total particle number constant. Hence, the distribution function deviates from a purely thermal distribution. The condensate fractions at the six time steps in the model calculation displayed in 
Fig.\,\ref{fig3} are 3, 10, 34, 49, 63, 74\,\%. Here, the last value refers to the stationary state, which is reached when the flow of particles into the condensate is equal to the diffusion out of it \cite{gw18a}, and the UV tail has acquired its thermal shape at $t=\tau^+_\text{eq}$.  For comparison, Boltzmann's distribution at  
the reduced temperature  $T^*\simeq 3T/4$ following the first evaporative cooling step is also shown, lower solid curve in Fig.\,\ref{fig3}.
\begin{figure}[t!]
\begin{center}
\includegraphics[width=7.6cm]{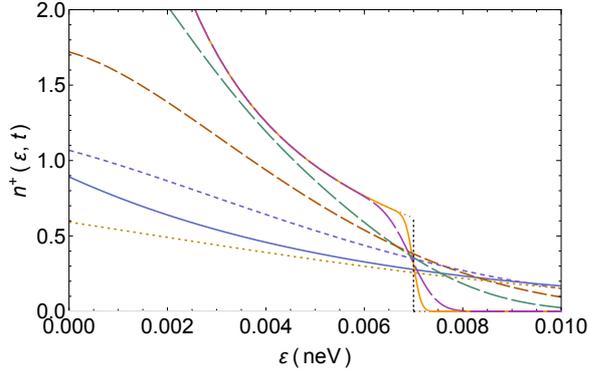}
\caption{\label{fig3}Occupation-number distribution $n^+(\epsilon,t)$ of a bosonic cold quantum gas based on the nonlinear evolution according 
to Eq.\,(\ref{bosfer})
	starting from a truncated equilibrium distribution Eq.\,(\ref{inibec}), upper curve with cutoff at $\epsilon_\text{\,c}$\,=\,7\,peV.
	The parameters are  $\mu= - 0.69 \, \text{peV}$,  $T=-D/\varv=8$\,peV\,$ \simeq93$\,nK, $\tau^+_\text{eq}=4D/(9\varv^2)\simeq3.6$\,ms.
The time sequence is displayed at $t = 0.001, 0.01, 0.2, 0.6, 1.5, 3.6$ ms (ordered by decreasing dash length). The lower solid curve is 
Boltzmann's distribution at the reduced temperature of $3T/4$.
}  
\end{center}
\end{figure}
\begin{figure}[t!]
\begin{center}
\includegraphics[width=7.6cm]{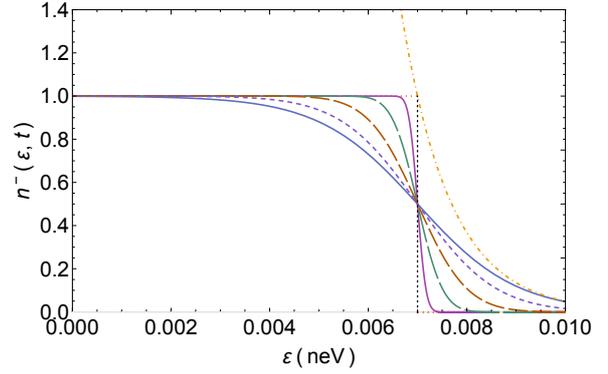}
\caption{\label{fig4}Occupation-number distribution $n^-(\epsilon,t)$ of a fermionic cold quantum gas based on the exact nonlinear evolution according to Eq.\,(\ref{bosfer}),
	starting from a $\theta$-function initial distribution (dotted) with $\epsilon_\text{\,F}=\mu=7$\,peV.
	The parameters are $T=-D/\varv$\,=\,1 peV\,$\simeq12$\,nK, $\tau^-_\text{eq}=4D/\varv^2\simeq7$\,ms.
The time sequence is displayed at $t = 0.01, 0.1, 0.5, 2$ ms (ordered by decreasing dash length towards the solid Fermi-Dirac equilibrium limit). The upper dot-dashed curve is the Boltzmann distribution.
}  
\end{center}
\end{figure}

In the above model calculation, the stationary value $N^\text{+\,eq}_\text{cond}(T)/N^\text{+}_\text{tot}$ of the condensate fraction turns out to be consistent with the established temperature dependence \cite{gro95,gps96} of the number of atoms in the condensed $p=0$ state 
\begin{equation}
\frac{N^\text{+\,eq}_\text{cond}(T)}{N_\text{tot}^+}=1-\left(\frac{T}{T_c}\right)^3\,.
\end{equation}

An example for the equilibration of a fermionic cold quantum gas such as $^6$Li according to the nonlinear diffusion equation starting from a theta-function initial distribution $\theta\,(1-\epsilon/\epsilon_\text{\,F})$ is shown in Fig.\,\ref{fig4}.
As in the corresponding high-energy case, the time-dependent analytical solutions approach the Fermi-Dirac equilibrium result. Boltzmann's distribution is given by the dot-dashed curve. The condensation of tightly bound fermion pairs \cite{zwi03} can, however, not be treated in the present approach.

\section{Conclusions}
To summarize, I have modelled the time evolution of locally equilibrating finite fermionic and bosonic systems through a nonlinear partial differential diffusion equation.
In the limit of constant transport coefficients, the equation is solved exactly, yielding in thermodynamic equilibrium the Fermi-Dirac distribution for fermions, and the Bose-Einstein distribution for bosons. The time-dependent solution of the nonlinear equation is well-suited to replace the linear relaxation-time approximation that has often been used in the literature.

For fermions, the time evolution including the production of antiparticles from the Dirac sea at high energies and temperatures is correctly described, such that overall particle-number and energy conservation are fulfilled. Using values of the equilibrium temperature and
the local equilibration time that are typical for the initial stages in central relativistic heavy-ion collisions at LHC energies, the transport coefficients and the time evolution of the valence-quark distribution, as well as of the distribution of gluons in the transverse degree of freedom have been calculated as an example of a relativistic system.
  
For bosons, the analytical modelling of the equilibration process through Eq.\,(\ref{bosfer}) is more challenging than in the fermionic case, because condensation may occur at sufficiently low temperatures, leading to a final state that differs from the purely thermal distribution \cite{svi91}. The buildup of the thermal tail for bosons in the ultraviolet, and the population of the condensate that is accounted for indirectly through the conservation of the total particle number has been demonstrated for a bosonic cold quantum gas. The time evolution of the thermal cloud in a fermionic cold quantum gas is described  as well using the analytical solutions of the nonlinear diffusion equation.\\

\noindent\bf{Acknowledgement}\rm\\

I am grateful to two of the referees for valuable suggestions.\\

\noindent\bf{References}\rm\\
\bibliographystyle{elsarticle-num}
\bibliography{gw_19}





\end{document}